\documentclass{article}

\usepackage{arxiv}

\usepackage[utf8]{inputenc} 
\usepackage[T1]{fontenc}    
\usepackage{hyperref}       
\usepackage{url}            
\usepackage{booktabs}       
\usepackage{amsfonts}       
\usepackage{nicefrac}       
\usepackage{microtype}      
\usepackage{lipsum}
\usepackage{graphicx}
\graphicspath{ {./images/} }
\title{Do We Need Explainable AI in Companies? Investigation of Challenges, Expectations, and Chances from Employees' Perspective}

\author{
 Katharina Weitz \\
 Chair for Human-Centered AI\\
 University of Augsburg\\
              Universit\"atsstraße 6a  \\
  86159 Augsburg \\
  \texttt{katharina.weitz@uni-a.de} \\
   \And
 Chi Tai Dang \\
 Chair for Human-Centered AI\\
  University of Augsburg\\
  Universit\"atsstraße 6a  \\
  86159 Augsburg \\
  \texttt{chitai.dang@uni-a.de} \\
\And
  Elisabeth Andr{\'e}\\
 Chair for Human-Centered AI\\
  University of Augsburg\\
  Universit\"atsstraße 6a  \\
  86159 Augsburg \\
  \texttt{elisabeth.andre@uni-a.de} \\
}

\begin{document}
\maketitle
\begin{abstract}
Companies' adoption of artificial intelligence (AI) is increasingly becoming an essential element of business success.
However, using AI poses new requirements for companies and their employees, including transparency and comprehensibility of AI systems.
The field of Explainable AI (XAI) aims to address these issues. Yet, the current research primarily consists of laboratory studies, and there is a need to improve the applicability of the findings to real-world situations. 
Therefore, this project report paper provides insights into employees' needs and attitudes towards (X)AI. For this, we investigate employees' perspectives on (X)AI.
Our findings suggest that AI and XAI are well-known terms perceived as important for employees.
This recognition is a critical first step for XAI to potentially drive successful usage of AI by providing comprehensible insights into AI technologies.
In a lessons-learned section, we discuss the open questions identified and suggest future research directions to develop human-centered XAI designs for companies.
By providing insights into employees' needs and attitudes towards (X)AI, our project report contributes to the development of XAI solutions that meet the requirements of companies and their employees, ultimately driving the successful adoption of AI technologies in the business context.

\keywords{Explainable AI \and Human-Centered AI \and User Evaluation}
\end{abstract}

\section{Introduction}
\label{intro}
AI applications have already impacted our private and work life, for example, voice and face recognition in smartphones, chatbots, or cobots in industry. National and international companies are aware of the impact of AI technology on their success and innovation potential. For example, European companies expect a sales increase with the help of AI of about 5 million US-Dollar until 2025 \cite{statista2016umsatzKI}. At the same time, legal regulations are demanding more and more that these AI technologies need to be comprehensible and transparent \cite{eu2018AIAct}. However, these requirements are not inherent, for example, in Deep Neural Networks (DNN), often referred to as \textit{black-box models}. 

The research area of Explainable AI (XAI) has dedicated itself to closing this gap and providing comprehensible explanations of black-box AI systems. 
Gunning et al. \cite{gunning2019xai} see the motivation of XAI in providing comprehensible explanations to humans. Therefore, DNNs become explainable when their inner workings or decisions are described so humans can understand them \cite{gilpin2018explaining2}. Besides this, researchers claim more and more that \textit{one explanation does not fit all}, and demand that XAI needs to be personalized, depending on the stakeholder and the application scenario \cite{bunt2012explanations,ehrlich2011taking}. 
The popular opinion is that XAI has different relevance for different stakeholders \cite{schneider2019personalized}. Therefore, Schneider and Handali \cite{schneider2019personalized} highlight the importance of collecting and investigating data and information of explainees (e.g., machine learning knowledge, preferences, prior experiences).

\begin{figure}
    \centering
    \includegraphics[width=0.8\linewidth]{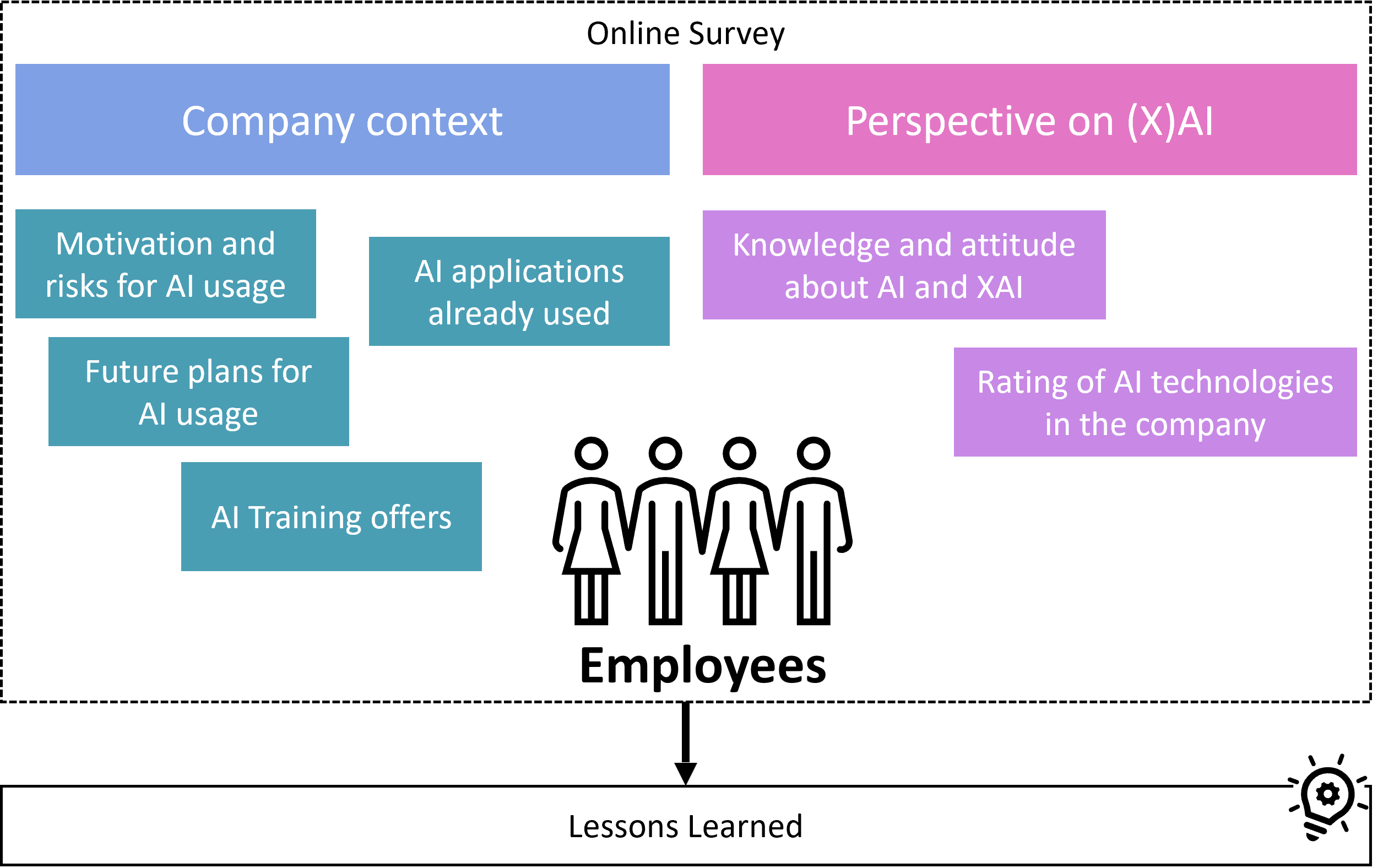}
    \caption{Overview of the content of this paper. 
    We describe the results of an online survey conducted in companies that asks employees about the \textit{company context} and their \textit{perspective on (X)AI}.
    Based on the results, open questions are discussed and future research directions to incorporate employees' requirements in XAI design are suggested.}
    \label{fig:ProcedureOverview}
\end{figure}

The overview of Langer et al. \cite{langer2021we} show that most of the scientific work in the field of XAI is done without empirical investigation of stakeholder. Langer et al. \cite{langer2021we} highlight the importance of investigating stakeholders' desiderata to develop and adjust XAI approaches. However, they also point out that little research currently identifies, defines, and empirically examines stakeholder desiderata.
Instead, the small amount of conducted experiments investigates the impact of explanations on users in different domains such as healthcare (e.g., \cite{mertes2021GANterfactual}), education (e.g., \cite{weitz2021demystifying}), and production work (e.g., \cite{rehse2019towards}). 
This research primarily focuses on laboratory experiments and gives first impressions and understandings about the impact of XAI on users referred to as \textit{human-grounded evaluation} \cite{doshi2017towards}.
The insights of these lab studies show that users' attributes and the AI application's characteristics must be considered when designing XAI \cite{weitz2021psychconcepts}. 
Nevertheless, studies need to pay more attention to the demands of real-world applications. Kraus et al. \cite{kraus2021KIkompetenz} investigated the impact of XAI in economically relevant ecosystems (e.g., healthcare, financial and manufacturing sector, construction industry). For this, they conducted an online survey and interviews with relevant stakeholders. The focus of their work lies primarily in the investigation of the helpfulness of different XAI tools for application-specific use cases. 
However, stakeholders' needs and attributes must also be considered when designing XAI \cite{holzinger2022personas}. Company employees' needs and attitudes regarding (X)AI must be clarified. Our study approaches this issue by asking employees about their attitudes towards (X)AI\footnote{(X)AI refers to AI and XAI} and their impressions of AI deployment in their company. 
Therefore, our paper investigates employees' perspective on (X)AI by using an online survey: (1) to understand the context of their work with AI, we investigated the employees' perspective of AI used in their company, including an overview of the current AI applications used, as well as (2) the personal perception of employees towards AI and XAI (see Figure \ref{fig:ProcedureOverview}). 

Based on the results, we discuss open questions in the lessons learned section. More concretely, this paper contributes the following:

\begin{itemize}
    \item It provides employees' views about their companies' usage of AI technology and their perception of XAI.
    \item It gives an overview of the challenges and risks of AI technology that employees perceive and provides impressions of users' needs.
    \item Finally, it concludes \textit{lessons learned} where we discuss open questions identified that serve as a basis to investigate requirements for human-centered XAI designs in companies and industries.
\end{itemize}

\section{Online Survey}

\subsection{Procedure}
Our goal was to identify the present
state of (X)AI-related issues and potential in companies from an employee's perspective. 
To achieve this, employees of companies of different sizes and sectors were asked about AI technology's current and future development in their company through an online survey. We distributed the questionnaire
through multipliers of the \textit{Plattform Lernende Systeme/acatech}\footnote{The National Academy of Science and Engineering (acatech) provides advice on strategic engineering and technology policy issues to policymakers and the public}
(e.g., chambers, competence centers, corporate leaders) to cover a broad portfolio of companies and their employees. The questionnaire was composed in German and aimed at employees of German-based companies.

\subsection{Research Questions}
We asked each employee about the current status and the strategic planning of using AI systems in their company. 
For this, we formulated the questions of the survey to address: 
(1) a broader view on AI in the \textit{company context} and (2) the \textit{perspective of employees} about (X)AI \footnote{Abbreviation interpretation of the research questions: RQ = research question, E = Employee, C = Company}. Since we want to investigate employees' perspectives as they interact with a (future) AI system in the company, it is important to note that the survey reflects the employees' subjective perception, not the company's slogan. 

\paragraph{Company Context}
The company perspective may generally serve companies that do not yet, hardly, or already use AI technologies for further strategic orientation and planning. For example, what are company motivations, usage areas, or issues regarding AI? Here, the experiences and decisions gained from the current state help assess the individual potential by introducing or using (X)AI technologies.
To understand the company and working context, we investigated the view of employees about AI in their company, including a look at the existing AI applications and those planned for the future. For this, we formulated the following research questions:
\begin{itemize}
    \item \textbf{RQ-C1}: What motivations and risks for their company do employees see in using AI technologies?
    \item \textbf{RQ-C2}: Do companies already use AI technology, and if so, which applications exist in companies?
    \item \textbf{RQ-C3}: What are companies' plans regarding using AI technologies?
\end{itemize}

\paragraph{Employees' (X)AI Perspective}
Insights from the general attitude, knowledge, or acceptance of (X)AI technologies, including demographics, from employees' perspectives show the state of practical implementation in companies. This guides the requirements regarding implementing (X)AI technologies. 
To investigate the personal perception of employees regarding (X)AI and their experiences with AI technologies in their companies, we formulated the following research questions: 
\begin{itemize}
    \item \textbf{RQ-E1}: How do employees rate their (X)AI knowledge and attitudes towards (X)AI? 
    \item \textbf{RQ-E2}: How do employees rate the AI technologies used in their company?
    \item \textbf{RQ-E3}: Is there a correlation between personal AI knowledge/attitude and the rating of AI technologies in the company?
    \item \textbf{RQ-E4}: How does the perception of the AI technology used in their company differ depending on demographic data (e.g., age, educational attainment, company position)?
    \item \textbf{RQ-E5}: How does the knowledge and attitude towards XAI relate to demographic data? 
\end{itemize}

\subsection{Methodology}
We derived a survey with groups of questions addressing each of our formulated research questions.

\paragraph{Demographic Data} We collected information regarding participants about their age, gender, educational background, knowledge, and role in the company.  

\paragraph{Company Information} To get an overview of the size and domain of the company, we asked questions about the sector and in which area (i.e., production or office work) the participants work. Here we used a combination of predefined answers and free-form answers. 

\paragraph{AI Technology - Strategy} To investigate the strategic plans towards AI for the company, we asked about plans for the usage of AI (e.g., ``In which areas do your company plan to make changes with the help of artificial intelligence in the next years?''). Furthermore, we addressed chances (i.e., ``What is driving AI development in your company?'') as well as risks (i.e., ``What are challenges, obstacles or problems for your company in implementing AI?''). We gave predefined answer options for each question and the possibility of writing free-text answers. 

\paragraph{AI Technology - Usage} Here, we requested detailed information about the 
AI technologies used (i.e., task/ goal of the AI, the field of application, the autonomy of the AI, duration of use). In addition, we asked, inspired by the overview of XAI metrics of Hoffman et al. \cite{hoffman2018MetricsXAI}, five items on a 7-point Likert scale (1 = not at all, 7 = extremely) regarding the AI technology's reliability, usefulness, transparency, operability, and comprehensibility. 

\paragraph{Training Offers} We investigated companies' general and AI-specific training offers. These investigations' analyses, results, and discussions were published in \cite{andre2021KIkompetenz}.

\paragraph{(X)AI Knowledge and Attitude} We investigated participants' knowledge and attitude about (X)AI. For knowledge, we asked whether they had heard the term AI and XAI. After that, we presented a definition of the terms. While AI is a broad area including a variety of different methods, we presented participants with a more general definition of AI: ``The term 'artificial intelligence' is often used to describe machines (or computers) that mimic 'cognitive' functions that humans associate with the human mind, such as ``learning'' and ``problem-solving''. This definition is oriented on the definitions given by Russell and Norvig \cite{russell2016artificial}. Due to an expected heterogeneous pool of company employees, we chose a broad definition of AI to verify that participants generally understood AI. In later questions, they had the chance to describe the specific AI systems used in their companies. XAI was described by highlighting its goal of it \cite{adadi2018peeking}: ``Explainable AI will enable people to understand, appropriately trust, and effectively manage AI technologies.''
Participants could agree or disagree with these definitions and provide their definition of (X)AI in a free-form text field. 
In addition to their knowledge, we asked participants about their attitude towards (X)AI on a 7-point Likert scale (1 = extremely negative to 7 = extremely positive). Finally, we collected their rating of the importance of XAI for different stakeholders (7-point Likert scale), 1 = I do not agree to 7 = I totally agree).

\subsection{Participants \& Companies}
\begin{figure}
    \centering
    \includegraphics[width=0.6\linewidth]{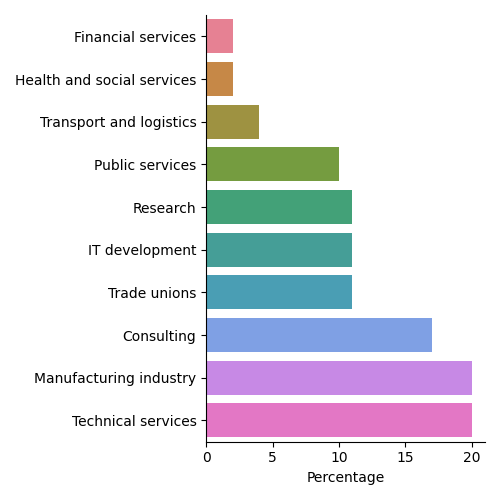}
    \caption{Participants represent a broad portfolio of companies. The Company sectors \textit{consulting}, \textit{manufacturing}, and \textit{technical services} were represented at most in our survey.}
    \label{fig:CompanySectors}
\end{figure}

\begin{figure}
    \centering
    \includegraphics[width=0.6\linewidth]{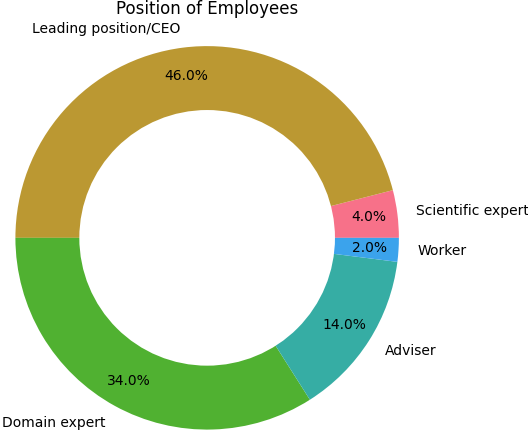}
    \caption{Most of the participants in our survey had a domain expert role or a leading position. Workers are barely represented in the survey.}
    \label{fig:EmployeesPosition}
\end{figure}

We collected data from 50 participants between 25 and 66 years (\textit{M}~=~45.0, \textit{SD}~=~11.3). Thirty-four of the participants were male, and 16 were female. 80\% of them had an academic educational background (i.e., bachelor's/master's degree or higher). 24\% were employed in medium-sized ($>$50 to $<$251 employees), 56\% in big-sized ($>$250 employees) companies. We received feedback from employees from a broad portfolio of companies (see Figure \ref{fig:CompanySectors}). Here, 80\% had a domain expert role, scientific expert role, or leading position. Workers and temporary staff were with 2\% barely represented (see Figure \ref{fig:EmployeesPosition}). 

\section{Results}
\subsection{Results for Company Context}

\textit{RQ-C1 to RQ-C3} The strongest \textit{motivation of companies} for using AI technology is an increase in productivity (\textit{n}~=~23), followed by an increase in flexibility (\textit{n}~=~21), customer requirements (\textit{n}~=~18), and adjustment of business models (\textit{n}~=~18). \textit{Risks} by using AI are financial aspects (\textit{n}~=~24), qualification of employees (\textit{n}~=~21), and acceptance by employees (\textit{n}~=~18). 
56.8\% of the participants stated that their company uses AI technology in prototypes (12 companies) or applications daily (13 companies). Furthermore, over two years, AI technology has been used in 54.2\% of companies. More details were revealed by the free-form answers about the \textit{application areas of AI}. Here, we found four clusters:

\begin{itemize}
    \item \textbf{Quality Assurance}: Mostly, participants stated that the
    AI technologies help monitor and predict production quality (e.g., by predictive maintenance using image classification), which assures the quality of the produced goods or the functioning of the machines used. 
    \item \textbf{Process Optimization}: Processes are optimised due to streamlining of processes (e.g., by automatically evaluating and clustering Big Data). This leads to a cost reduction due to shorter and more efficient processes.
    \item \textbf{Support Employees}: AI is also used to support employees in fulfilling their tasks successfully, especially in office work. The usage of AI here covers a broad spectrum, from a simplification of bookkeeping to support of office-based work processes (e.g., software as a service\footnote{Buxmann et al. \cite[p.500]{buxmann2008software} describe the usage of the software as a service:``customers are provided with a standard software solution as a service via the Internet.''})
    \item \textbf{Interaction \& Communication}: This includes communication with customers or employees using chatbots (e.g., check-in process of a guest in a hotel) as well as interaction in the form of robots within a physical environment (e.g., intelligent positioning where a robot pick up goods). 
\end{itemize}

For the \textit{future}, participants stated that their companies focus on the usage of AI to change processes within the organization (\textit{n}~=~29), followed by the goal of developing new technologies (\textit{n}~=~25), and changes in the organization of the company (\textit{n}~=~21). 
Although we have a small sample of employees, the clusters found for AI applications (e.g., process optimization), as well as the risks (e.g., qualification of employees) seen in the use of AI in companies, are very similar to the results of larger surveys from over 500 industry companies in Germany \cite{statista2019hemmnisseKI,statista2020vorteileKI}.

\subsection{Results of Employees' perspective on (X)AI}

\paragraph{RQ-E1} All of the participants have heard of the term ``artificial intelligence``. 87\% of them agreed with our given definition of AI. Five participants had a different definition of AI in mind, especially focusing on ``the cloud`` rather than on physical machines or indicating that the term ``AI imitates human behavior'' is not correct to them. 
62.5\% of the participants had heard about the term XAI, while 37.5\% did not. Participants found XAI relevant for all the interest groups queried (items ranged from 1 = not important to 7 = very important), especially for companies (\textit{M}~=~6.05, \textit{SD}~=~1.41) and politicians (\textit{M}~=~5.90, \textit{SD}~=~1.50). 

Conducting a one-sample t-test, we found that participants had a significantly positive view towards AI compared to the mean value of the rating scale (i.e., \textit{M}~=~4, 7-point Likert scale), \textit{t}(39)~=~7.92, \textit{p}~=~$<$~.001, \textit{d}~=~1.25 (large effect)\footnote{The effect size \textit{d} is calculated according to Cohen \cite{cohen2013statistical}. Interpretation of the effect size is: \textit{d}~$<$~.5~:~small effect; \textit{d}~=~0.5-0.8~:~medium~effect; \textit{d}~$>$~0.8~:~large~effect}.

\begin{figure}
    \centering
    \includegraphics[width=0.6\linewidth]{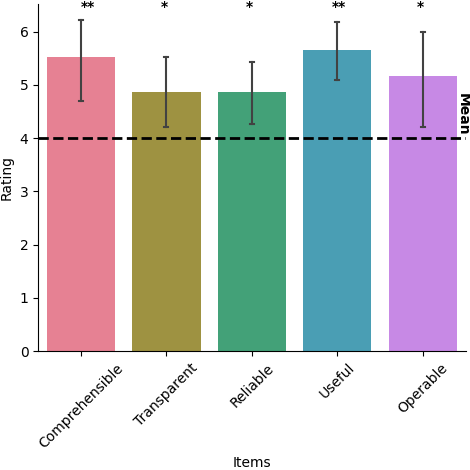}
    \caption{Rating of the AI technologies used in companies by employees. Employees perceived the AI technology significantly positively, compared to the mean of the rating scale, *\textit{p}~$<$~.05, **\textit{p}~$<$~.001. Error bars represent the 95\% CI.}
    \label{fig:AIRating}
\end{figure}

\paragraph{RQ-E2} Employees rated their experience with AI technology in the company significantly positive compared to the mean of the rating scale (i.e., \textit{M}~=~4, 7-point Likert scale) (see Table \ref{tab:ratingAItechnology}) for the items comprehensibility, transparency, reliability, usefulness, and operability (see Figure \ref{fig:AIRating}).

\begin{table}
    \caption{Rating of AI technology used in companies on five items. A one-sample t-test revealed that employees perceived all items as significantly positive.}
    \label{tab:ratingAItechnology} 
    \centering
    \begin{tabular}{lcrc}
    \hline\noalign{\smallskip}
    \textbf{Rating item} & \textit{t}(22) & \textit{p} & \textit{d} \\
    \noalign{\smallskip}\hline\noalign{\smallskip}
    useful & 5.79 & $<.001$** & 1.21 \\
    reliable & 2.87 & .009* & 0.60 \\
    operable & 2.60 & $.016$*~ & 0.54 \\
    comprehensible & 3.88 & $<.001$** & 0.81\\
    transparent & 2.43 & $.024$** & 0.51\\
    \noalign{\smallskip}\hline
    *\textit{p}~$<$~.05, **\textit{p}$<$.001
    \end{tabular}
\end{table}

\paragraph{RQ-E3} We found a significant positive correlation\footnote{We calculated Spearmans' Rang correlations} between employees' attitude towards AI and their rating of the AI technology in their company, \textit{$r_{sp}$}~$=$~.71, \textit{p}~$<$~.001, meaning that the higher the personal attitude towards AI of the employees, the higher is their positive perception of the AI technology in their company. The same significant positive relationship was found for the employees' attitude towards XAI and their rating on their company's AI technology, \textit{$r_{sp}$}~$=$~.56, \textit{p}~$=$~.007. 

\paragraph{RQ-E4 and RQ-E5} Demographic attributes such as age, gender, and educational background of employees did not correlate with the perception of AI technology in their companies, but their \textit{role in the company} did \textit{$r_{sp}$}~$=$~.-61, \textit{p}~$=$~.003.
This indicates that participants with a higher position in the company perceived the AI technology as less favourable. 

The knowledge about XAI correlates positively with educational background, \textit{$r_{sp}$}~$=$~.53, \textit{p}~$<$~.001.
Regarding XAI attitude, the demographic attributes \textit{company role}, \textit{$r_{sp}$}~$=$~-.42, \textit{p}~$=$~.008, and \textit{educational background}, \textit{$r_{sp}$}~$=$~.38, \textit{p}~$=$~.015 showed a significant correlation.
These correlations indicate, similar to the perception of AI in the company that a higher company position leads to a less positive attitude towards AI. However, at the same time, the educational background seems to positively impact the general knowledge and attitude towards XAI. 

Although the correlations found are interesting, they solely indicate a relationship, not a causal link. Further studies are needed to investigate the direction of the impact in more detail.
In the lessons-learned section, we will explore recommendations for further research based on the correlation findings.

\section{Lessons Learned}
\begin{figure}
    \centering
    \includegraphics[width=0.8\linewidth]{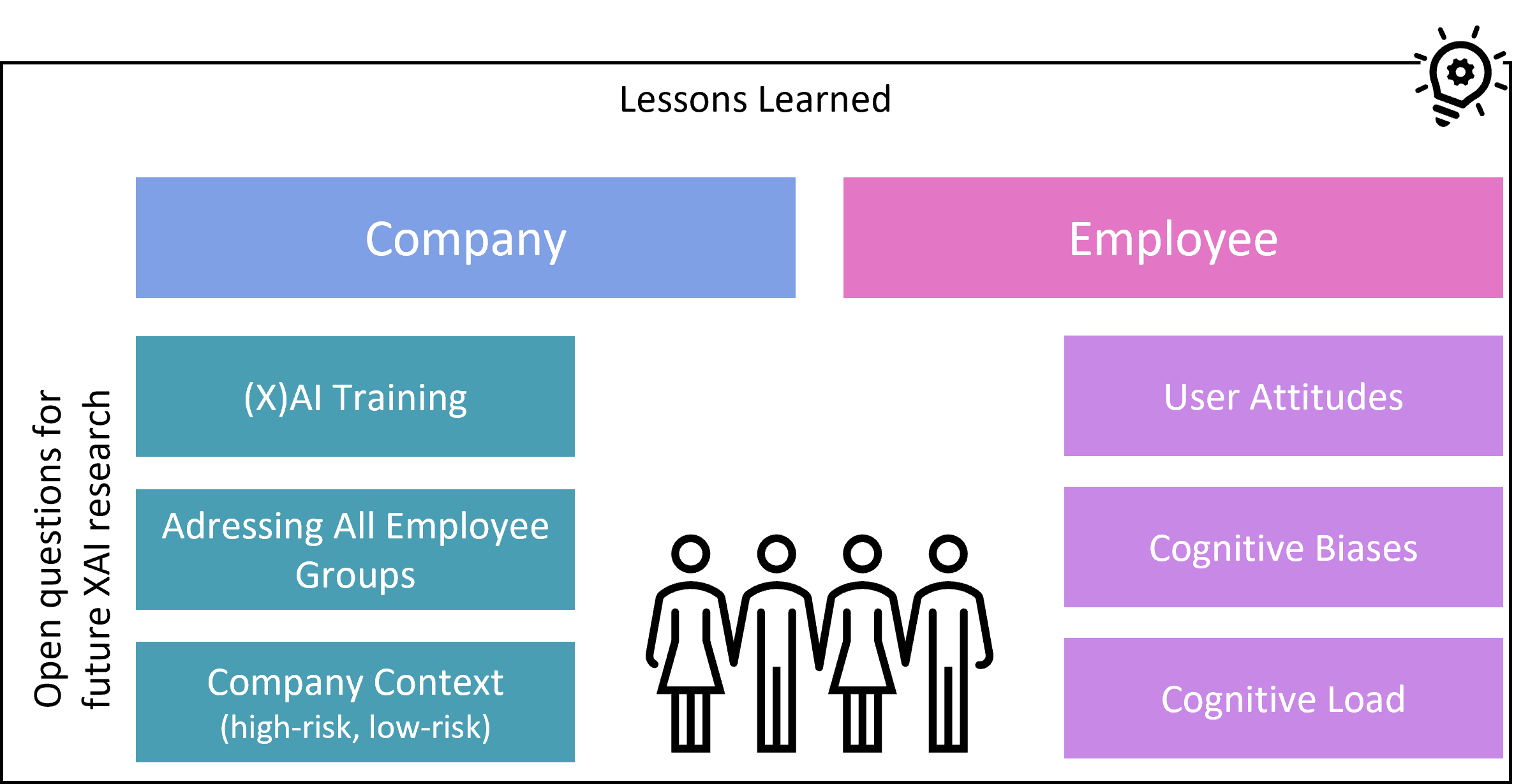}
    \caption{Overview about open research questions regarding XAI in companies we identified with our survey.}
    \label{fig:LessonsLearnedOverview}
\end{figure}

Based on the results of our online survey, we report lessons learned addressing requirements form an employee perspective that should be considered when designing and evaluating XAI for companies (see Figure \ref{fig:LessonsLearnedOverview}).

\subsection{Investigate the Impact of Cognitive Biases for (X)AI} We found a positive correlation between employees' attitude towards AI and their rating of AI technology used in the company. 
This supports the interpretation that there is a relation between the personal attitude towards AI and the perception of AI in the company context. The fact that we found a positive correlation here could indicate a cognitive bias, i.e., the \textit{halo effect}. The halo effect describes that attributes of a person or an object in one situation can positively or negatively impact the perception of this person/object in another situation \cite{thorndike1920constant}. For example, when employees perceive AI positively in one area, they also perceive it positively in another. 

\textbf{Suggestion:} Cognitive Biases such as the halo effect should be investigated in future (X)AI studies. For example, it could indicate that people with a negative opinion towards AI transfer this attitude to using AI in their companies if users' mental models are confirmed here (e.g., AI works unreliably). On the other hand, helpful AI technology could contribute to improving users' AI attitudes in another area. Whether these directions of effects exist and how they are related could be the subject of future studies. 

\subsection{Investigate the Influence of Company Context on XAI Requirements}
Our survey participants included employees from various sectors, predominantly from the manufacturing industry and technical services.
With this variability of companies comes a variability of tasks that AI is supposed to fulfil in these companies. For example, businesses that handle high-risk tasks have different requirements for XAI than businesses that operate more low-risk tasks. At the same time, explanations in the future will have to comply with stricter standards depending on the risk potential of the AI application (e.g., \cite{eu2018AIAct}). In addition to the influence of risk on XAI in the company context, other variables are also relevant to the design of XAI. For example, activities that must be carried out under time pressure may have different requirements for explanations (brief, succinct,  not interrupting production flow). In contrast, those for development purposes may require the most comprehensive information possible.

\textbf{Suggestion:} 
Future research should take into account the company-specific characteristics and requirements and their impact on the design of XAI.

\subsection{Investigate the Influence of User Attitudes on (X)AI Usage}
For AI, we found no correlations between employees' attitudes towards AI with demographic data except for a negative correlation with their company role.
For XAI, we found that the knowledge and the attitude about XAI are positively correlated to the educational background while it was negatively correlated to the company position of the employees. 
Similar to our findings, Weitz et al. \cite{weitz2021demystifying} found that demographic information such as age and gender do not impact users' perception of (X)AI in an educational setting. In contrast, a higher educational background positively impacted participants' trustworthiness in the AI system.

\textbf{Suggestion:} Hence, it is highly worthwhile to create and foster a positive attitude towards AI from the beginning, especially in the leading management, to achieve appropriate trust\footnote{Appropriate trust refers to trust in a technical system that matches its true capabilities \cite{lee2004trust}} and successful usage of deployed AI technologies later. Lichtenthaler \cite{lichtenthaler2019extremes} highlights the challenge to foster positive attributes towards AI in employees. He suggests assigning employees as ``gatekeepers'' to support the usage of AI in the company by positively influencing other employees. However, supporting appropriate trust in AI by XAI that also highlights the faults and errors of AI could be challenging, for that solutions should be investigated in more detail in the future.

\subsection{Explore the Effect of XAI on Cognitive Load} Our results suggest that our respondents perceive AI technology as already comprehensible and transparent. XAI is a known term for many employees, contrary to earlier studies' findings (e.g., \cite{heimerl2020unraveling}). This indicates that company employees are more in touch with the problem of explainability and are aware that this is an important topic. 
As reflected in the ratings, XAI is considered necessary, especially for companies. 
This awareness represents a fruitful basis for developing XAI for real-world applications. Nevertheless, based on the same survey data, we found in \cite{andre2021KIkompetenz} 
that AI usage leads to increasing requirements of employees. 

\textbf{Suggestion:} When XAI is perceived as important and valuable, can it help to support employees to handle the requirements for AI, or does it even increase the cognitive load in employees? While the works of \cite{abdul2020cogam, hudon2021explainable} investigate the impact of XAI on cognitive load in lab settings, an investigation in real-world scenarios, such as in a company is missing and could be a valuable investigation for future research.

\subsection{Companies Should Foster (X)AI Training} 
While respondents stated that they know about XAI, it remains unclear whether the individuals have a relatively shallow understanding of XAI or whether they have in-depth knowledge. To gain in-depth knowledge, training is needed to support employees in interacting efficiently and effectively with (X)AI.
However, training employees regarding AI is seen as a challenge for companies, as we reported in \cite{andre2021KIkompetenz}. 

\textbf{Suggestion:} Future studies should investigate whether this knowledge about XAI is shallow or whether users already have in-depth knowledge about XAI. Do users know different XAI methods? Do they use them in their work? How can companies benefit from the usage of XAI?
We can imagine two possibilities for beneficial XAI usage in companies: 
(1) XAI to support employees in their tasks. 
Here, XAI aims to provide good explanations supporting people's work (e.g., diagnosis of malfunctioning parts). (2) XAI can be used to train employees, for example, to explain the inner workings of AI (in training) to help employees work successfully with it by understanding it better, i.e., gaining AI competence \cite{long2020ai}. In addition, XAI can help reduce fears towards AI technology that employees may have.
To identify beneficial use cases for XAI in companies, further studies have to investigate which and to what extent XAI methods are or should be used in the company.

\subsection{XAI Research Needs to Address All Employee Groups} Generally, we found that employees perceive the company's AI technologies as comprehensible, transparent, reliable, useful, and operable. While these results are encouraging, it is essential to note that we have responses almost exclusively from male employees between 39-39 years with academic backgrounds who are leaders or have domain expertise. Therefore, whether employees with other backgrounds have similar impressions remains to be seen.  
In addition, by recruiting participants via the \textit{Plattform Lernende Systeme/acatech}, a selection bias \cite{heckman1979sample}, leading to responses, especially from people interested in the topic and therefore having a more positive view towards (X)AI could be possible. 

\textbf{Suggestion:} Thus, for further studies on XAI in companies, special attention should be paid to reaching other target groups, such as workers and untrained staff who operate with AI. It should also be ensured to investigate not only technology-interested or technology-accepting employees but a wide range of employees' opinions. A safe, data-protected, trustworthy connection with the employees must be established for this.

\section{Conclusion}
The integration of AI into corporate strategic plans has the potential to enhance company success and innovation, making it a key area of focus for businesses.
However, legal regulations have compelled companies to prioritize comprehensibility in their AI systems. XAI refers to methods that address this challenge.
Although initial research on the impact of XAI on end-users is emerging from lab experiments, real-world applications remain understudied. 
Therefore, it is crucial to consider the perceptions and needs of employees when designing and evaluating XAI for companies in a human-centered way.
Our online survey has moved research closer to this goal by investigating employees' perspectives towards X(AI). Our findings in this project report suggest that
XAI is already a known topic among employees and is perceived as an important issue. 
Based on our results, we suggest (1) investigating in more detail the impact of cognitive biases and user attributes on (X)AI usage, (2) investigating the role of the company context on requirements for XAI design (3) exploring the effect of XAI on cognitive load (4) fostering (X)AI training in companies, and (5) considering all employee groups in XAI research.
Our insights encourage researchers to include employees' attitudes towards (X)AI in their design to create a more human-centered XAI.

\section*{Acknowledgment}
We thank our partners from the university of Stuttgart, Fraunhofer IAO, and Plattform Lernende Systeme/acatech who worked with us on the micro-project, as published in \cite{andre2021KIkompetenz}.

\bibliographystyle{unsrt}  
\bibliography{literature.bib}

\end{document}